\documentclass{PoS}

\pdfoutput=1
\usepackage{cite}
\usepackage{amsmath}
\usepackage{amsfonts}
\usepackage{amssymb}
\usepackage{latexsym}
\usepackage{multirow}
\usepackage{color}
\usepackage{slashed}
\definecolor{rossos}{cmyk}{0,1,1,0.55}
\definecolor{bluscuro}{rgb}{0.15, 0.2, .85}
\definecolor{bluchiaro}{cmyk}{1,.3,0.,0.1}

\definecolor{rossos}{cmyk}{0,1,1,0.55}
\definecolor{bluscuro}{rgb}{0.15, 0.2, .85}
\definecolor{bluchiaro}{cmyk}{1,.3,0.,0.1}

\newcommand{\bc}{\begin{center}}
\newcommand{\ec}{\end{center}}
\newcommand{\pMET}{{\bf p}\llap{/\kern1.5pt}_T}
\newcommand{\bea}{\begin{eqnarray}}
\newcommand{\eea}{\end{eqnarray}}
\newcommand{\ignore}[1]{}
\newcommand{\be}{\begin{equation}}
\newcommand{\ee}{\end{equation}}

\def\l{\label}

\def\({\left(}
\def\){\right)}
\def\<{\langle}
\def\>{\rangle}
\def\f{\frac}
\def\be{\begin{equation}}
\def\ee{\end{equation}}
\def\bry{\begin{array}}
\def\ery{\end{array}}
\def\bes{\begin{subequations}}
\def\ees{\end{subequations}}
\def\bit{\begin{itemize}}
\def\eit{\end{itemize}}
\def\ben{\begin{enumerate}}
\def\een{\end{enumerate}}
\def\dst{\displaystyle}

\newcommand{\MET}{E\llap{/\kern1.5pt}_T}
\definecolor{grey}{rgb}{0.6,0.6,0.6}
\definecolor{fuchsia}{rgb}{1,0,1}
\newcommand{\red}[1]{{\color{magenta}\color{red}#1\color{magenta}}}

\title{Light stop squarks and b-tagging}

\ShortTitle{Light stop squarks and b-tagging}

\author{Gabriele Ferretti\\
	Department of Fundamental Physics, Chalmers University of Technology, 412 96 G\"oteborg, Sweden\\
	E-mail: \email{ferretti@chalmers.se}}
	
\author{Roberto Franceschini\\
	Theory Division, Physics Department, CERN, CH-1211 Geneva 23, Switzerland\\
	E-mail: \email{roberto.franceschini@cern.ch}}
	
\author{Christoffer Petersson\\
	Physique Th\'eorique et Math\'ematique, Universit\'e Libre de Bruxelles, C.P. 231, 1050 Brussels, Belgium\\
	International Solvay Institutes, Brussels, Belgium\\
	Department of Fundamental Physics, Chalmers University of Technology, 412 96 G\"oteborg, Sweden\\
	E-mail: \email{christoffer.petersson@ulb.ac.be}}

\author{\speaker{Riccardo Torre}\\
        Dipartimento di Fisica e Astronomia, Universit\'a di Padova and\\
        INFN, Sezione di Padova, via Marzolo 8, I-35131 Padova, Italy\\
        E-mail: \email{riccardo.torre@pd.infn.it}}

\abstract{A significant part of the parameter space for light stop squarks still remains unconstrained by collider searches. For both R-Parity Conserving (RPC) and R-Parity Violating (RPV) scenarios there are regions in which the stop mass is around or below the top quark mass that are particularly challenging experimentally. Here we review the status of light stop searches, both in RPC and RPV scenarios. We also propose strategies, generally based on exploiting $b$-tagging, to cover the unconstrained regions.}

\FullConference{To appear in the Proceedings of the Corfu Summer Institute 2014 "School and Workshops on Elementary Particle Physics and Gravity", Corfu, Greece
{\vspace{-20.5cm}\hspace{3cm}\flushright{\small \textnormal{DFPD-2015-TH-12\\} }\vspace{20.5cm}}}

\begin{document}

\section{Introduction}\l{sec:Introduction}
Supersymmetry (SUSY) is one of the best frameworks to address and solve the hierarchy problem, i.e.~to stabilize the Higgs potential against radiative corrections. The existence of new particles, superpartners of the Standard Model (SM) ones, around or just above the weak scale is the most striking prediction of supersymmetric theories, as far as Naturalness is concerned. Experimental searches at colliders have put significant pressure on SUSY, due to the lack of any signal of physics beyond the SM. This translates into a high level of fine-tuning, often already below the percent level, especially for the minimal realization of SUSY, the Minimal Supersymmetric Standard Model (MSSM). Beyond the MSSM, constraints can be made milder, often at the price of some theoretical complication. And even in those scenarios, due to the general lack of signals of colored particles below the TeV scale, some tension with Naturalness typically remains.


However, relevant regions of the parameter space of SUSY theories with light colored superpartners, such as the stop squarks, are still unconstrained by collider searches. This is mainly due  to the particularly challenging final states they produce. In R-Parity Conserving (RPC) SUSY, compressed spectra typically give rise to very ``soft'' particles and a small amount of missing transverse energy $\MET$, and often lead to top-like final states which are difficult to disentangle from the $t\bar{t}$ SM process. In (baryonic) R-Parity Violating (RPV) SUSY, each stop decays to two quarks, leading to multi-jet final states that suffer from the very large QCD background, which makes searches, especially at low masses, particularly difficult. Light stops have been widely studied in the recent literature by both the theoretical \cite{Plehn:2010we,Bornhauser:2010gb,Brust:2011gf,Kats:2011fk,Bi:2011jv,BinHe:2011tw,Drees:2012hw,Bai:2012ux,Plehn:2012mz,Alves:2012gf,Han:2012ve,Kaplan:2012wf,Brust:2012yq,Ghosh:2012cx,Choudhury:2012hg,Evans:2012oq,Kilic:2012yq,Graesser:2012dn,Krizka:2012ii,Franceschini:2012vl,Delgado:2012eu,Dutta:2013gd,Buckley:2013wo,Chakraborty:2013iu,Low:2013kl,Bai:2013wx,Belanger:2013wy,Boughezal:2013cn,Han:2013ui,Dutta:2013wy,Papucci:2014ue,Buckley:2014bf,Czakon:2014wa,Grober:2014vl,Eifert:2014uu,Cho:2014vd,Beuria:2015ua,Batell:2015wg,Rolbiecki:2015wb,Padley:2015wx,Hikasa:2015vw} and experimental \cite{Aad:2013ija,Aad:2014qaa,ATLAScollaboration:2014kf,Aad:2014bva,Aad:2014kra,ATLAS:2013aia,CMScollaboration:2013dj,Chatrchyan:2013xna,CMS-PAS-SUS-13-009,CMScollaboration:2013gy,CMScollaboration:2014jo,CMScollaboration:2014tk,CMScollaboration:2014wb,CMS:2014wsa} communities.

To test such unconstrained regions of the SUSY parameter space is clearly relevant from the point of view of Naturalness. It may be argued that, due to the generally stronger constraints on gluinos (often above \mbox{1 TeV}), the relevance of Naturalness for light stops is weakened and that more theoretical ingredients are needed to obtain a consistent picture. However, from a pure phenomenological point of view, probing stop squarks with masses around and below the top quark mass is still of great importance, regardless of any theoretical prejudice.

In this proceeding we review the status of light stops both in the case of RPC SUSY and in the presence of baryonic RPV interactions. In both cases we stress the importance of triggers and $b$-tagging in probing the unconstrained regions.

The document is organized as follows. In Section \ref{sec:RPCreview} we review the status of stop squark searches after the first LHC run and in Section \ref{sec:RPCnew} we propose a new monojet-like analysis, exploiting $b$-tagging, to test four body stop decays in the compressed region. Section \ref{sec:RPVreview} is devoted to review the state of light stop squarks in the presence of baryonic RPV interactions, when the stops decay to a pair of quarks. In Section \ref{sec:RPVnew} we propose a strategy, again based on $b$-tagging, to cover the gap between previous collider and existing LHC searches. Conclusions are drawn in Section \ref{sec:Conclusion}.

\section{Light stop searches in RPC SUSY}\l{sec:RPCreview}



In the RPC case, the simplified model we consider  involves the lightest of the two stop squarks, $\tilde{t}_1$, which decays to the lightest supersymmetric particle (LSP), the neutralino $\tilde{\chi}^{0}_{1}$. 
We distinguish three main kinematical regions characterized by the stop-neutralino mass difference $\Delta m=m_{\tilde{t}_1} - m_{\tilde{\chi}^{0}_{1}}$. In the case when the mass difference is larger than the top mass, $\Delta m > m_t $, the stop always decays promptly via the two-body decay to a top quark and the neutralino. 
This is the region where the strongest bounds on the stop mass are attained~\cite{CMS:2014wsa,Aad:2014kra} -- for a neutralino lighter than 250\,GeV, stop masses below 600--750\,GeV are excluded. Up until recently, there was a unconstrained triangular region for an almost massless neutralino and 180 $\mathrm{GeV} < m_{\tilde{t}_1} < 200~\mathrm{GeV}$,  but this triangle has been essentially closed by the constraint on the stop production cross section coming from the spin correlation measurement of $t\bar{t}$ production \cite{ATLAScollaboration:2014vi}.

In the intermediate region $m_W + m_b < \Delta m < m_t $, the two body decay is precluded and the stop decays via a three body process involving the $W$-boson and the $b$-quark from the off-shell top, and the neutralino~\cite{Chatrchyan:2013xna,Aad:2014kra,Aad:2014qaa}. Also in this region the stop decay is prompt. 
The current data exclude stop masses up to roughly 300\,GeV in the central region of this domain.

Lastly, for $0 < \Delta m < m_W + m_b$, the above channels are closed and the stop decays to the neutralino and either one or three light SM fermions. The first option corresponds to the one-loop decay process $\tilde{t}_1\to c + \tilde{\chi}^{0}_{1}$, 
while the second option is the logical extension of the decays discussed above, where now even the $W$ is forced to be off-shell, i.e.~$\tilde{t}_1\to b + f + f' + \tilde{\chi}^{0}_{1} $. This is perhaps the most difficult region to investigate since there are two competing decay channels with model dependent branching ratios. The model dependency is mainly due to the first process, which involves the masses of all the superpartners that enter the loop as well as the squark flavor structure. Moreover, in the kinematical region $0 < \Delta m \lesssim 20\,\mathrm{GeV}$, the partial life-time of the second decay becomes larger than
$0.1 \mathrm{mm}/c$, giving rise to displaced vertices or R-hadrons, unless the first decay option dominates and is prompt. Thus, different search strategies must be employed in order to cover all these cases.

A survey of the current bounds in the $m_{\tilde{t}_1}, m_{\tilde{\chi}^{0}_{1}}$ plane is presented in Figure~\ref{SummaryPlots} (left). The majority of searches in the most squeezed region have targeted the decay mode $\tilde{t}_1\to c + \tilde{\chi}^{0}_{1}$, assuming this decay mode to have 100\% BR~\cite{ATLAS:2013aia,CMS-PAS-SUS-13-009}. Under this assumption, stop masses below 250 GeV have been excluded.  The four body decay process $\tilde{t}_1\to b + f + f' + \tilde{\chi}^{0}_{1} $ has been targeted by the ATLAS searches~\cite{Aad:2014kra,ATLAS:2013aia}, in which 100\% BR is assumed. However, under these assumptions, there is still a fairly large unconstrained region at low masses. Let us again stress that, for the extremely squeezed case, $0 < \Delta m \lesssim 20\,\mathrm{GeV}$, this decay is unlikely to be prompt. For this decay mode, the only model dependence comes from the interaction vertex between stop-top-neutralino, all other interactions and masses being well known in the SM. Such a vertex cannot deviate too much from the typical electroweak strength and thus the width for this process is dictated by the four-body phase-space.
Yet, in the presentation of the analysis, the experimental collaborations assume the decay to be prompt throughout the entire region and present their exclusion limits all the way until they reach the line $\Delta m = m_b$. Of course, one may argue that this assumption is made in the spirit of simplified models, avoiding any theory bias and thus considering the stop-neutralino model as a paradigm for a more generic simplified model of a colored scalar decaying into an invisible fermion and three additional light fermions. As a cautionary remark though, our proposed improvement on the search strategy for this channel \cite{Ferretti:2015uy} will heavily rely on one of these ``light'' fermions being a $b$-quark thus using some of the theoretical ingredients from the MSSM.

The two searches~\cite{ATLAS:2013aia,CMS-PAS-SUS-13-009} introduced in the above discussion cover complementary kinematical regions.
The search~\cite{ATLAS:2013aia} is basically a monojet search in which at least one hard jet is required in order to have a sufficient recoil that gives rise to a large amount of missing transverse energy ($\MET$). The very nature of the selection criteria is such that the search is most sensitive near the $\Delta m = m_b$ region where the stop decay products are soft. On the contrary, the search~\cite{CMS-PAS-SUS-13-009} is most sensitive near the region $\Delta m =  m_W + m_b $. This is so because the search requires the presence of a lepton in the final state that needs to be sufficiently hard to be reconstructed. 
Due to the difficulties in the reconstruction of soft leptons using fast detector simulation,  it is in general difficult to recast this search. Similarly, the proposal of its improvement suffers from the difficulty of dealing with soft leptons. For this reason we do not attempt to recast and improve this latter search. Instead we have chosen to focus on improving the former search strategy~\cite{ATLAS:2013aia}.

We are now ready to discuss the targeted region of our proposal. In Figure~\ref{SummaryPlots} (left) we summarize the status of searches for light stops in the squeezed region with the assumption of a 100\% BR into a four-body final state. There is an unconstrained region of approximately triangular shape for $90~\mathrm{GeV}\lesssim m_{\tilde{t}_1} \lesssim  140\mathrm{~GeV}$\,GeV, bounded by the exclusion curves from the ATLAS \mbox{1-lepton} search \cite{Aad:2014kra}, the ATLAS monojet search \cite{ATLAS:2013aia} and LEP \cite{Heister:2002hp}. We have decided to enlarge the
search region to include lower values for the stop mass, down to $m_{\tilde{t}_1} = 80\mathrm{~GeV}$. Hence, we include smaller stop masses than the usual region discussed in the ATLAS summary plot that terminates at $m_{\tilde{t}_1} = 110\mathrm{~GeV}$. The reason is that this region is still unconstrained by any direct search and, since the lightest stop may well be hiding there, we should allow for this possibility. Note that, in some models, such a light stop may give rise to a problematic contribution to the Higgs production cross section and precision observables but, since these contributions are more model dependent and can in some cases be canceled by additional degrees of freedom, we prefer to take the agnostic approach and allow for the full parameter space.

Before moving to the proposed improvement in the search strategy, we would like to comment on the production modes for the stop~\cite{Ferrettietal:2015boh}. Clearly ordinary stop pair production via strong interactions has the largest cross section throughout the parameter space. There are however a couple of additional production modes that might be of interest. 
In the simplified model we are considering, there is still an unconstrained  region where the top quark is heavier than the \emph{sum} $m_{\tilde{t}_1} + m_{\tilde{\chi}^{0}_{1}}$. In this region the top can develop an exotic decay mode $t\to \tilde{t}_1 + \tilde{\chi}^{0}_{1}$ in addition to the ordinary SM decay mode. The branching ratio for this exotic decay is in principle constrained by top physics measurements, but the visible final states after the stop decays are the same, masking the signal, and thus a non-negligible branching fraction could be possible. 
However in the context of the processes and final states we are considering, this production mode does not significantly alter the results and we will ignore it in the following. 
 We remark that the amount of top decays $t\to \tilde{t}_1 + \tilde{\chi}^{0}_{1}$ is a function of the sum  of masses $m_{\tilde{t}_1} + m_{\tilde{\chi}_{1}^{0}}$, that is  independent from  $m_{\tilde{t}_1} - m_{\tilde{\chi}_{1}^{0}}$, which rules the branching ratios into  2-, 3- and 4- body final states of the stop decay. Therefore the investigation of top decays into stop might prove very useful to put robust bounds on light stops {\it without} a strong dependence on the stop-neutralino mass difference, which instead is a nuisance of the searches for direct QCD production of stops.

A second possible production mode is the 2-to-3 hard process $p p \to t \tilde{t}_1 + \tilde{\chi}^{0}_{1}$ that is present even if both $t\to \tilde{t}_1 + \tilde{\chi}^{0}_{1}$ and $\tilde{t}_1 \to t + \tilde{\chi}^{0}_{1}$ are kinematically forbidden. Being a 2$\to$3 process with three massive particle in the final state, its cross-section is greatly suppressed by the phase-space factor. However this process has the advantage of not requiring the additional jet to recoil against, since the two neutralino are not produced back to back as in the usual $2\to 2$ pair production case. In this way, the 2$\to$3 process disentangles the usual relation between $\Delta m$ and the amount of $\MET$. 

\section{Monojet with $b$-tags}\l{sec:RPCnew}

In \cite{ATLAS:2013aia} ATLAS performed the search for pair produced compressed stop squarks that our proposal is trying to improve upon. They considered a total of five signal regions. Two of them (C1 and C2) target the decay mode $\tilde{t}_1 \to c + \tilde{\chi}^{0}_{1}$ and will not be considered further. The remaining three (M1, M2 and M3) are monojet-like searches that target both the charm decay mode and the four-body decay mode we are interested in. 
It is perhaps a bit misleading to call such search ``monojet'' given 
the fact that the actual selection cut employed is $N_{\mathrm{jets}}\leq 3$. At any rate this search should not be confused with the ``Dark Matter'' monojet search in~\cite{ATLAS-Collaboration:2014xia}.

After a preselection, that includes a lepton veto, the three ATLAS monojet regions are first characterized by a common set of selections criteria, namely the presence of at most three jets with $p_T > 30\mathrm{~GeV}$ and $|\eta| < 2.8$ and an azimuthal angular separation between these jets and the $\MET$, $\Delta \phi > 0.4$. The difference between the three regions is then based on the $p_T$ of the leading jet and the $\MET$ in the event. The values for these last two cuts are $(p_T^{j_1},\MET)=(280,220), (340,340)$ and (450,450) GeV in the signal regions M1, M2 and M3 respectively.
The different signal regions are optimized for different regions of parameter space. Since we are interested in targeting the low-mass region we only consider the  M1  selections in the following. Our proposal is quite straightforward and can be summarized in one line: \emph{To improve the sensitivity to the low mass region, add a $b$-tag requirement on one of the (at most) three jets.}

More specifically, we require the presence of at least
one $b$-tagged jet with \sloppy\mbox{$ 30\mathrm{~GeV} < p_T < 300\mathrm{~GeV}$}, $|\eta| < 2.5$.
In the above $p_T$ and $\eta$ range, the ATLAS calibration algorithm for $b$-tagging is data-driven, thus reducing the systematic uncertainties coming
from Monte Carlo simulations. In what follows we denote the new signal region by ``M1+$b$-tag".

Looking at the background estimations by ATLAS for the M1 signal region it is easy to see why the addition of a $b$-tag is expected to improve the sensitivity to the signal. For $ 20.3 \mathrm{fb}^{-1}$ at 8~TeV, out of a total of expected 33450 $\pm$ 960 background events, 17400 $\pm$ 720 come from SM processes involving $Z \to \nu\nu$ and $14100\pm 337$ come from the processes involving $W \to \ell\nu$. The key point is that both of these leading backgrounds are dramatically reduced by the extra $b$-tag requirement. 

We simulated both signal and background by using {\sc MadGraph5} \cite{Alwall:2014vb}, {\sc Pythia6}
\cite{2006JHEP...05..026S}, \sloppy\mbox{{\sc FastJet3}} \cite{Cacciari:2011rt,Cacciari:2006vn} and {\sc Delphes3} \cite{deFavereau:2013fe}. In the fast detector simulation we used the standard ATLAS detector specification.
The PDF set used is the CTEQ6L1, jets are reconstructed using the anti-$k_t$ algorithm \cite{Cacciari:2008hb} with $\Delta R = 0.4$ and MLM matching \cite{MLM1,Mangano:2006cp} is used throughout the simulation. Table~\ref{table:BGevents} summarizes the expected leading backgrounds with and without the $b$-tagging requirement.

\begin{table*}[t!]
\small\begin{center}
\begin{tabular}{l|cccccc}
Background				&$t\bar{t}$			&$Z({\to}\nu\nu)$	&$W({\to}\ell\nu)$		&Dibosons		&Others		&Total \\ \hline\hline
M1 (ATLAS~\cite{ATLAS:2013aia})		&$780\pm 73$
	&$17400\pm 720$	&$14100\pm 337$		&$650\pm 99$
	& $565\pm301$	&$33450\pm 960$\\ 	
M1+$b$-tag	 			&$307\pm 57\ignore{\red{(35)}}$	&$261\pm 22$		&$144\pm 7$			&$55\pm 17\ignore{\red{(13)}}$	& -	&$767\pm 64$\\  \hline
\end{tabular}
\end{center}\vspace{-2mm}
\caption{\small\label{table:BGevents} Estimated numbers of background events with 20.3\,fb${}^{-1}$ of 8\,TeV LHC data.
The background is given as $B \pm \delta B$, $B$ being the central value and $\delta B$ the $1 \sigma$ error. The error in the M1 case is simply taken from ATLAS. For the error in the M1+$b$-tag region we quote twice the relative error, see \cite{Ferretti:2015uy} for discussion.
}\vspace{-2mm}
\end{table*}

In order to reproduce as faithfully as possible the ATLAS situation we used the following strategy.
We generated, fully independently, a large sample of background events and passed them through the same cuts as those performed by ATLAS in M1.
From this analysis, we found the central values for all the backgrounds to be within 20\% of the ATLAS results. This gives us confidence that the remaining background sample is representative of the physical situation after the cuts. Since we are only interested in the further improvements in efficiency arising from the $b$-tagging, we normalize the expected number of events to the ATLAS numbers and only multiply by the $b$-tag efficiency we get when passing from M1 to M1+$b$-tag.

Table~\ref{table:BGevents} conveys the idea of the discriminating power of the $b$-tag in rejecting invisible $Z$ decays and leptonic $W$ decays. The two leading backgrounds in~\cite{ATLAS:2013aia} are now subleading with respect to the $t \bar t$ background which, since it contains $b$-jets, is only mildly affected by the extra cut. The signal, as will be discussed below, is found to behave in a similar way as the $t \bar t$ background and thus the overall sensitivity is significantly improved.

Turning now to the signal, we simulated a grid of points with $ 70~\mathrm{~GeV} \leq m_{\tilde t_1} \leq 250~\mathrm{~GeV}$ and
$0~\mathrm{~GeV} \leq m_{\tilde \chi^0_1} \leq 200~\mathrm{~GeV}$ in steps of 10~GeV inside the region $10~\mathrm{GeV} \leq \Delta m \leq 80\mathrm{~GeV}$. As mentioned above, we only considered stop pair production as production mode.
For stop masses above 100 GeV we used the NLO+NLL cross sections of \mbox{ATLAS \cite{LHCSUSY}} while for the few points below \mbox{100 GeV}, we used {\sc Prospino} \cite{Beenakker:1998fr} to determine the main slope and fixed the absolute normalization with the ATLAS values above 100\,GeV.

\begin{figure*}[t!]
\begin{center}
\hspace{-4mm} \includegraphics[scale=0.35]{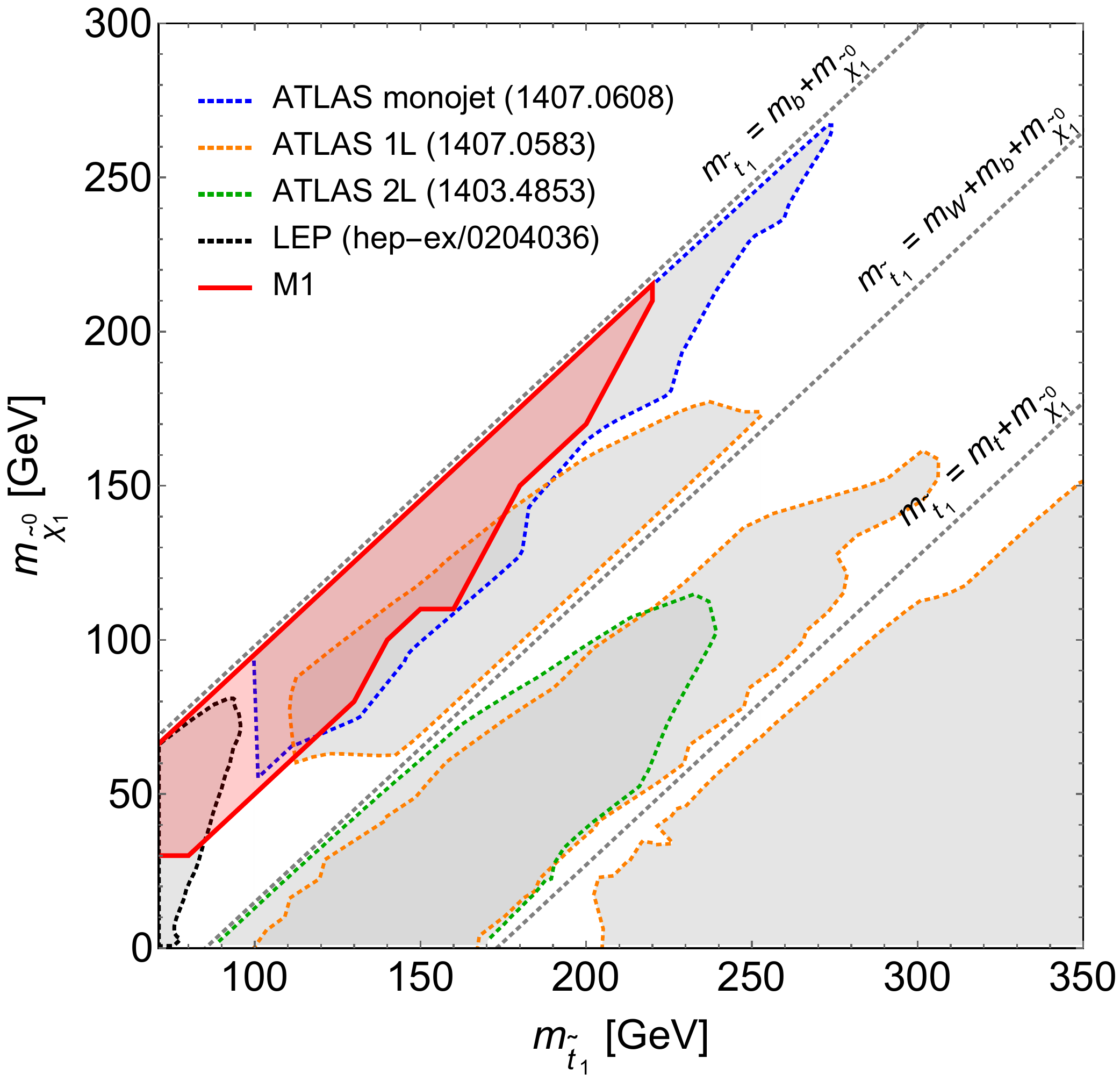}\hspace{8mm}
\hspace{-4mm} \includegraphics[scale=0.35]{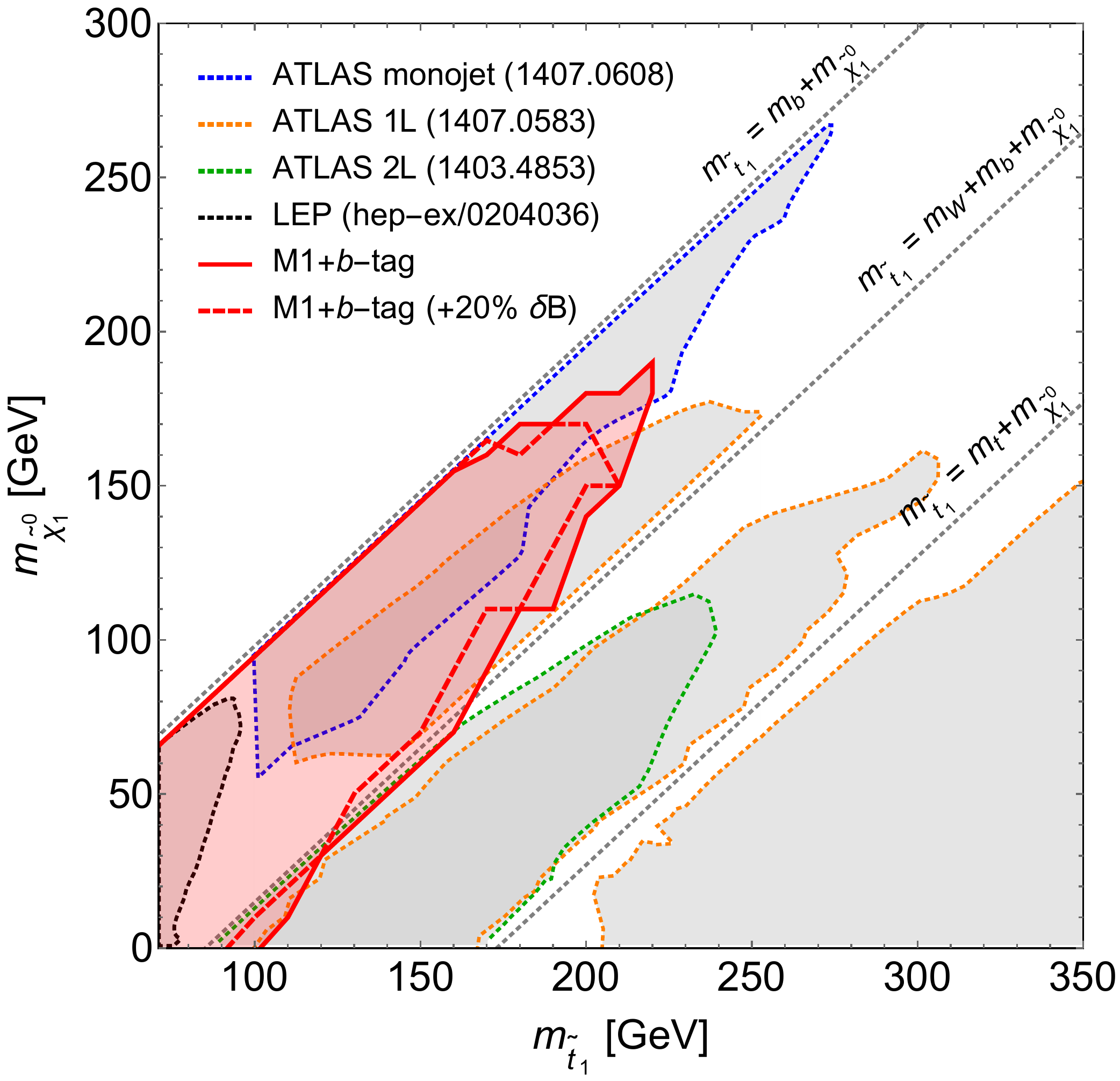}
\caption{Left: Existing limits in the stop-neutralino mass plane, superimposed with the M1 exclusion curve as obtained in our analysis and extended to
$m_{\tilde{t}_1} > 100\mathrm{~GeV}$. Right: The main result from our proposed search ``M1+$b$-tag", (red solid curve), with the change induced by increasing the background error by 20 \% (red dashed curve). Figure taken from ref.~\cite{Ferretti:2015uy}.}\vspace{-4mm}
\label{SummaryPlots}
\end{center}
\end{figure*}

After imposing the M1 cuts we again found agreement within 20\% with ATLAS. We used the same normalization procedure as for the background, normalizing the number of events before the $b$-tag requirement to the ATLAS numbers and using only the $b$-tag efficiency to obtain the final results for the signal.
The main difference with the previous background simulation is that, due to the large number of points on the grid and the small efficiencies, we are unable to generate a statistically significant fully matched sample of events. We thus resort to the following strategy. For each point we generate two exclusive samples, one containing zero jets at the parton level and one containing exactly one such jet with $p_T > 200$~GeV.
The ratio of the LO cross sections obtained is used to estimate the efficiency of the $p_T$ cut:
$\epsilon_{p_T > 200\mathrm{~GeV}} = \sigma(pp\to \tilde{t}_1 \tilde{t}_1 j(p_T>200\mathrm{~GeV}))/\sigma(pp\to \tilde{t}_1 \tilde{t}_1)$. The one-jet unmatched sample is then used throughout the analysis. Checking this procedure on a limited number of points we found good agreement with the results form the fully matched sample. The results are presented in Figure~\ref{SummaryPlots} (right).

We set limits by excluding points for which the number of events $N > 1.96\delta B$. Notice that the analysis is essentially all driven by systematics and not by statistics. In the left panel of Figure~\ref{SummaryPlots} we present, for comparison, the exclusion limits we obtain repeating the M1 analysis of ATLAS (red exclusion curve). We are able to reproduce their exclusion boundaries fairly accurately, which validates the analysis on the right. Note that the limits in~\cite{ATLAS:2013aia} are broader since they take into account the signal regions M2 and M3 that become more relevant for large masses. We have tested imposing the $b$-tag requirement on those regions as well but the low signal efficiency makes them not viable, at least for the $ 20.3 \mathrm{fb}^{-1}$ 8~TeV data. In fact, we tested changing the $p_T$ and $\MET$ cuts within the intervals defined by M1 and M2 and we found that M1 essentially gives the best $S/B$ ratio within statistical fluctuations. 
The red curve in the right panel of Figure~\ref{SummaryPlots} is our main result. It fully covers the so far unconstrained region exposed in the left panel of Figure~\ref{SummaryPlots} and even extends slightly into the three body region.

With an eye to the upcoming Run-2 of LHC it is worth studying if a search of this type can be further optimized to cover an even larger region. We already mentioned that the increased energy and luminosity will allow for stronger cuts and perhaps even studies of different production modes with lower cross sections. Here we conclude by pointing out that it may be possible also to move in the opposite direction and \emph{relax} some of the constraints imposed by the current search. The M1 ATLAS cut flow was heavily relying on the isolation requirement between the jets and the $\MET$, \mbox{$\Delta \phi > 0.4$}. This is necessary to reduce the QCD jet contamination and $\MET$ from jet mismeasurements. 
The requirement of a $b$-jet however already dramatically cuts the QCD background and could be considered an alternative to the former requirement. However, due to the uncertainties involved in calculating the QCD background with our simulation tools, we refrain from making any estimates of the potential gain in sensitivity. Nevertheless, we would like to encourage the experimental collaborations to also consider this possibility.

\section{Light stop searches in RPV SUSY}\l{sec:RPVreview}
As discussed in the previous sections, the lack of signals of SUSY so far can be due to a compressed spectrum that leads to final states with low $\MET$ and soft SM particles. Alternatively, SUSY can lead to final states that do not contain large $\MET$ for more structural reasons. The most interesting such situation arises when the assumption of conservation of R-parity, usually made in the context of minimal SUSY models, is relaxed. Without R-parity, when extending the SM to a supersymmetric theory, the following superpotential terms arise:
\be
W_{\text{RPV}}=\mu'_{i}L_{i} H_{u}+\f{1}{2}\lambda_{ijk} L_{i}L_{j}E^{c}_{k}+\lambda'_{ijk} L_{i}Q_{j}d^{c}_{k}+\f{1}{2}\lambda''_{ijk}u^{c}_{i}d^{c}_{j}d^{c}_{k}\,.
\ee
The interactions proportional to $\mu',\lambda,\lambda'$ break lepton number $L$ (lRPV), while the one proportional to $\lambda''$ breaks baryon number $B$ (bRPV). These two classes of interactions cannot be present together with un-suppressed couplings since they would induce too fast proton decay. Moreover, lepton number violating ($\slashed{L}$) interactions are typically more constrained by collider searches than baryon number violating ($\slashed{B}$) ones, because they usually generate signals with many leptons in the final state. Weaker constraints instead apply to $\slashed{B}$ interactions. Couplings involving the first two generations are subject to strong constraints coming from a number of low energy processes such as $n-\bar{n}$ oscillations, di-nucleon decays, $K-\bar{K}$ mixing.
A comprehensive report of these constraints is given in Refs.~\cite{Barbier:2005rr,DiLuzio:2013wh} and is summarized in Table \ref{table:RPVconstraints}.
\begin{table*}[t!]
\begin{center}
\begin{tabular}{l|l}
Process		& Constraint \\ \hline\hline
$NN\to K^{+}K^{+}$	& $|\lambda_{uds}''|<O(10^{-5})$ \\ \hline
$n-\bar{n}$			& $|\lambda_{udb}''|<O(10^{-3})$ \\ \hline
$n-\bar{n}$			& $|\lambda_{tds}''|<O(10^{-1})$ \\ \hline
$n-\bar{n}$			& $|\lambda_{tdb}''|<O(10^{-1})$ \\ \hline\hline
$K-\bar{K}$			& $|\lambda_{cdb}''\lambda_{csb}''|<O(10^{-3})$ \\ \hline
$K-\bar{K}$			& $|\lambda_{tdb}''\lambda_{tsb}''|<O(10^{-3})$ \\ \hline
$B^{+}\to K^{0}\pi^{+}$	& $|\lambda_{ids}''\lambda_{idb}''|<O(10^{-1})$ \\ \hline
$B^{-}\to\phi\pi^{-}$		& $|\lambda_{ids}''\lambda_{isb}''|<O(10^{-3})$ \\ \hline\hline
\end{tabular}
\end{center}\vspace{-2mm}
\caption{\small\label{table:RPVconstraints} Summary of constraints on the $\slashed{B}$ couplings $\lambda''$ for superpartner masses $\widetilde{m}=500$ GeV \cite{DiLuzio:2013wh}.
}\vspace{-2mm}
\end{table*}
One immediately sees from the table that the constraints on the $\lambda''$ couplings involving the third generation quarks are typically  mildly bound, while there are strong bounds on the couplings involving the first two generation quarks. This suggests two things: On one hand, it makes the search for third generation bRPV squarks (stop and sbottom squarks) at the LHC particularly interesting. On the other hand it suggests that an hierarchical pattern of $\lambda''$ couplings is necessary to expect a signal while still respecting the constraints in Table \ref{table:RPVconstraints}. This is indeed the case in most of the theoretical constructions aimed at explaining the pattern of bRPV couplings. In particular, several different approaches suggest rather similar hierarchical structures for the $\lambda''$ couplings, like for instance Minimal Flavor Violation (MFV) \cite{Nikolidakis:2008tu,Csaki:2012zr}, Partial Compositeness (PC) \cite{Keren-Zur:2012cr}, unification \cite{DiLuzio:2013wh} and dynamical RPV \cite{Csaki:2013vs}. Most of these frameworks generally predict $\lambda''$ couplings of the form
\be\l{RPVcouplings}
\lambda_{ijk}''\propto V_{il}^{\text{CKM}}\(\f{m_{u_{i}}m_{d_{j}}m_{d_{k}}}{m_{t}^{3}}\)^{\mu}\epsilon_{ljk}
\ee
with $\mu=1$ (for different constructions with $\mu<1$ see Ref.~\cite{Franceschini:2013tm}). The structure in eq.~\eqref{RPVcouplings} with $\mu=1$ in particular implies $\text{BR}\(\tilde{t}_1\to bd+bs\)\approx 99\%$, which means that bRPV decays of the stop squark typically contain a $b$-quark in the final state.

Until recently, bRPV stops were only excluded by LEP and Tevatron searches for two resonances in four-jet final states. The most constraining bounds from LEP come from the OPAL Collaboration \cite{OPALCollaboration:2003hu} and give
\be
m_{\tilde{t}_1}\(\theta_{\tilde{t}}=0.98\)\geq 77\text{ GeV}\,\qquad \text{and}\qquad  m_{\tilde{t}_1}\(\theta_{\tilde{t}}=0\)\geq 88\text{ GeV}\,,
\ee
with $\theta_{\tilde{t}}$ being the stop squark mixing angle, while the CDF Collaboration has set the bound \cite{CDFCollaboration:2013up}
\be
m_{\tilde{t}_1}\leq 50 \text{ GeV}\wedge m_{\tilde{t}_1}\geq 100 \text{ GeV}\,,
\ee
independent of the stop mixing angle.

 It should be remarked that hadron collider experiments have difficulties to probe very light stops, in the case of CDF in fact there is no sensitivity to masses below 50~GeV. This difficulty arises because of the need to trigger on hard experimental objects, such as jets of hadrons, in such experiments. For too small stop mass the energy of the final state particles is simply not enough to trigger the detectors. However, thanks to the combination with LEP bounds, stop squarks decaying through bRPV interactions with a mass smaller than 100~GeV are excluded.

  Recently, the CMS Collaboration, considering both decays to four jets and to two jets and two $b$-quarks has obtained the bounds \cite{CMScollaboration:2014wb}
\be
\bry{l}
\dst m_{\tilde{t}_1}\leq 200 \text{ GeV}\wedge m_{\tilde{t}_1}\geq 350 \text{ GeV}\,,\qquad \text{for}\qquad \text{BR}\(\tilde{t}_1\to jj\)=1\,,\\
\dst m_{\tilde{t}_1}\leq 200 \text{ GeV}\wedge m_{\tilde{t}_1}\geq 385 \text{ GeV}\,,\qquad \text{for}\qquad \text{BR}\(\tilde{t}_1\to bj\)=1\,.
\ery
\label{cmsJJbJ}
\ee
The region of stop masses between $100$ GeV and $200$ GeV remains the only unconstrained one below $350$ GeV. Also in this case the reason that makes this region particularly challenging experimentally is related to the minimum trigger requirements that the LHC has to employ to tag events with four jets. The minimum $p_{T}$ that can be achieved by these triggers is typically bigger than about $80$ to $100$ GeV which makes the analyses almost completely insensitive to final states with less than $400$ GeV of invariant mass, and therefore to stop masses below $200$ GeV. In the next section we discuss how the region between $100$ and $200$ GeV, particularly motivated by naturalness arguments, could be covered provided events with suitable triggers have been recorded or will be available in the future.

\section{Paired di-jet searches with $b$-tags}\l{sec:RPVnew}

Despite the recent progress from the results of the CMS collaboration, it is discomforting to find that stops of mass $100 \textrm{ GeV} < m_{\tilde{t}_{1}} < 200 \textrm{ GeV}$ are not yet directly excluded. As explained, this is the result of the too high trigger thresholds under which the experiments must operate to avoid triggering on the unbearably huge amount of QCD events producing multi-jet final states. This problem is in fact common to other searches for multi-jet resonances, such as the search for gluino  production and its RPV decay $\tilde{g}\to jjj$ for which, in the region around $m_{\tilde{g}}\simeq 140\textrm{ GeV}$, resonance searches have not been able to put a bound~\footnote{In this case, however, a bound from pure counting of multi-jet final state events exists \cite{ATLAScollaboration:2012dx}. While this bound seems to exclude cross-sections far below the gluino production cross-sections, it is important to obtain an independent bound from resonance searches, which are less subject to the large uncertainties that affect QCD predictions for pure rates.}.

An ideal solution to this lack of sensitivity for light colored matter produced at the LHC would be to attempt its direct search in low instantaneous luminosity data, as has been done in Ref.~\cite{Aad:2011ly}. This search could exploit low trigger thresholds used in the 2010 LHC 7 TeV run. However the limited integrated luminosity of just 34~pb$^{-1}$ was not sufficient to be sensitive to the production cross-section of stops.

In absence of a plan for a low instantaneous luminosity run of the LHC with low trigger thresholds, other strategies have been put forwards to attempt the search of light stops.
A simple idea to combat the QCD background and lower trigger thresholds has to do with the presence of heavy flavors in the decay of stops. Unlike the case of RPC supersymmetry, where the flavor structure of the CKM mixing matrix guarantees a $b$-quark in the final state, the flavor of the quarks into which the stop decays is depending on the unknown RPV couplings. Despite this, it has been argued that it is very likely that a relation between the SM fermions Yukawa and the bRPV coupling exists in model that explain dynamically the origin of the RPV couplings. The upshot of these model building activities~\cite{Nikolidakis:2008tu,Csaki:2012zr,Keren-Zur:2012cr,DiLuzio:2013wh,Franceschini:2013tm,Csaki:2013vs} is that
most of the times a stop decays into a final state with at least one heavy flavor quark $\tilde{t}_1 \to bj$. For this reason, in the case of RPV stops it seems particularly motivated to search for traces of heavy flavor quarks in multi-jet final states.
This idea has been pursued in \cite{Franceschini:2012vl} where it was shown that the simple requirement of two $b$-tags in the multi-jet final state can reduce the background by 1 or 2 orders of magnitude, depending on the chosen $b$-tagging algorithm, with just a mild reduction of the signal rate. The improvement in signal-over-background rate is particularly evident in the search of lighter stops. However, even for larger stop masses the advantage persists, as shown by the extended range of stop masses (\ref{cmsJJbJ}) that CMS has excluded.

Ref.~\cite{Franceschini:2012vl} shows that the sensitivity to light stops could also be improved employing more targeted kinematic selections, aimed at preserving as much as possible signals with jets of low transverse momentum and pairs of low invariant masses.

A first point that has been highlighted is that, when searching for light stops, the partitioning of the four jets into two pairs, each corresponding to the candidate stop resonance, is best done using angular correlations of the jets~\cite{Schumann:2011fj}, rather than
the invariant mass of the candidate stops resonances. Despite the latter option might seem more intuitive, the use of angular correlations produces candidate stop mass distributions that are smoother, in particular at low masses. Therefore the pairing of jets according to angular criteria seems to offer advantages in searching for light stops.

A second important point made in Ref.~\cite{Franceschini:2012vl} is that the shape of the background, hence our ability to isolate a signal, depends quite sensitively on {\it intra}-resonance jets angles. In practice the jets from each candidate stop resonance tends to be more or less collinear. It is useful to make the selection $$|\delta\eta_{ab}|+|\delta\eta_{cd}| < \delta\eta_{intrares} \quad\textrm{ and }\quad |\delta R_{ab}|+|\delta R_{cd}| < \delta R_{intrares}$$ where $ab$ and $cd$ indicate the two pairs of jets forming the candidate stop resonance and $\delta \eta_{intrares}$ and $\delta R_{intrares}$ are two values (around 1.5 and 3) to be optimized to smoothen the shape of the background in the region of candidate stop mass that one wants to probe.

The idea to use the angles between the two jets arising from each stop, and in particular to require a certain degree of collinearity between them, is highly suggestive of the production of boosted stop resonances, which might result in hadrons that are merged in single jets by the jet clustering algorithms. In fact a search strategy for light boosted stops has been proposed in \cite{Bai:2013ta}, which finds sensitivity to light stops in the currently not excluded region of stop masses already in the 8~TeV data set of the LHC.

As seen in Table~\ref{table:RPVconstraints} and in the related discussion, the RPV couplings are in general expected to be small. In fact their magnitude is often so small to imply that a particle that can decay only through RPV couplings might have a detectably long life-time, above 0.1mm$/c$. In this case the existence of tracks from non-prompt decays can be used to identify signal events. Despite many hadronic resonances having non-prompt decays, hence giving rise to a background for non-prompt RPV stop decays, the presence of non-prompt tracks, and especially of displaced secondary decay vertexes in the event, can be used to put rather severe bounds on light RPV stops with life-time between 0.1~mm$/c$ and 100~m$/c$ \cite{Liu:2015wx}. Furthermore, since the stops are colored particles, if they do not decay promptly they will form hadrons. Some of these hadrons will have electric charge and searches for (meta-)stable charged objects would put bounds on such RPV stop hadrons  for life-times  between 1~m$/c$ and 1~Km$/c$ \cite{Liu:2015wx}.

\section{Conclusion}\l{sec:Conclusion}
In this contribution we discussed searches for light stop squarks
, both in RPC and RPV scenarios. Despite the extensive experimental program to constrain all the allowed regions for light stop squarks, unconstrained regions still remain, with stop masses around or below the top quark one. We found that, by using suitable analysis techniques based on exploiting the presence of one or two $b$-quarks in the final state, such light stops could be completely covered, already by using the  8 TeV LHC dataset\footnote{In the case of RPV stops this essentially depends on whether suitable trigger configurations were used during data taking.}. 
This is expected to have important implications for the idea of Naturalness.

\section*{Acknowledgments}
The work of C.\,P.~is supported by the Swedish Research Council (VR) under the contract 637-2013-475, by IISN-Belgium (conventions 4.4511.06, 4.4505.86 and 4.4514.08) and by the ``Communaut\'e Fran\c{c}aise de Belgique" through the ARC program and by a ``Mandat d'Impulsion Scientifique" of the F.R.S.-FNRS.
The work of R.T. was supported by the ERC Advanced Grant no.~267985 {\it DaMeSyFla} and by the Italian PRIN no.~2010YJ2NYW$\_$003. We also acknowledge the grant SNF Sinergia no.~CRSII2-141847.

\bibliographystyle{mine}
\bibliography{bibliography}

\end{document}